# Quantum Markov Chains: Hub-Pruned Estimation for Fashion Recommenders

Or Peretz[a], Tai Dinh[b], Michal Koren[a]

[a] Shenkar—Engineering. Design. Art, Ramat-Gan, 5252626, Israel
[b] The Kyoto College of Graduate Studies for Informatics, Sakyo, Kyoto City, Kyoto, 606-8225, Japan

**Abstract**
We investigate whether shallow quantum circuits can accurately reproduce the short-horizon dynamics of discrete-time Markov chains derived from fashion electronic-commerce recommendation links. Transition operators are compiled into block-encoded circuits and iterated using fixed-point oblivious amplitude amplification, and amplitude-encoded marginals are used to estimate the classical push-forward. Empirically, colour categories such as black and, to a lesser extent, white function as high-degree hubs that dominate probability flow. Consequently, we assess three chain variants: the full network including all colours, a network without black, and a network without black and white, to quantify the effect of hub pruning under realistic circuit depths and measurement budgets. Across networks aggregated from multiple retailers, hub pruning consistently improves quantum and classical agreement at shallow depth; total-variation distance and Kullback-Leibler divergence typically decrease by approximately a factor of two relative to the full network, while state fidelities remain close to unity. A bias-and-contraction analysis explains these gains through reduced cross-terms and an effectively widened spectral gap. The results identify hub-pruned block-encodings as a practical heuristic for near-term experiments on recommendation dynamics using small quantum registers, and they provide a reproducible benchmarking protocol that reports total-variation distance, Kullback-Leibler divergence, and state fidelity as functions of circuit depth and prediction horizon.

## 1. Introduction

Discrete-time Markov chains (DTMCs) over observable colour states provide a compact lens on short-horizon navigation and preference dynamics in fashion e-commerce. Platform-exposed recommendation links induce transitions between product colours; aggregating these edges yields a row-stochastic operator whose powers summarize multi-step flows across the colour palette. In practice, the induced graphs are highly non-uniform: a small set of hub colours, most prominently black and, to a lesser extent, white, concentrate probability mass and short-circuit medium-probability relations that are arguably more diagnostic for ranking diversity and exploration. This paper examines whether shallow quantum compilations of such DTMCs can recover the relevant distributional features with high fidelity, and how hub structure mediates estimation quality under realistic depth and shot budgets.

Classical estimators remain the default for transition inference and stationary analyses, but they incur sampling burdens for small-mass events and, in hubbed graphs, can bias early transients toward degree-dominated paths. Quantum primitives promise complementary advantages.

Amplitude encoding converts distributions to amplitudes so that overlaps and divergences become geometric quantities. Amplitude estimation (AE) reduces Monte-Carlo precision scaling from $O\left(\frac{1}{\varepsilon^2}\right)$ to $O\left(\frac{1}{\varepsilon}\right)$, and block-encodings and quantum signal processing transform stochastic operators at low polynomial degree (Szegedy, 2004; Montanaro, 2015; Low & Chuang, 2017; Gilyén et al., 2019). Yet these asymptotics are conditional: state preparation, oracle design, and finite depth can erase nominal speedups, and skewed success amplitudes magnify finite-shot variance in the tails (Suzuki et al., 2020; Giurgica-Tiron et al., 2020). The central question is therefore not whether quantum methods can help in principle, but when shallow, practically compiled circuits match or exceed classical precision in regard to the metrics that matter for colour-aware recommendation.

We address this question by compiling colour-transition DTMCs into block-encoded updates with oblivious amplitude amplification, and by comparing quantum vs. classical marginals and divergences across three chains: the full catalogue, the catalogue without black, and the catalogue without black and white. The design isolates how hub dominance affects conditioning and, consequently, shallow-depth agreement. We evaluate agreement between quantum and classical estimators using total variation distance (TVD), Euclidean distance, Kullback-Leibler divergence (KL), and state fidelity. Our contributions are as follows: (1) Model and compilation: we formalize colour-level DTMCs and construct depth-aware quantum updates via block-encoding and fixed-point amplification, with explicit handling of padding and induced subgraphs for hub-pruned variants (Low & Chuang, 2017; Yoder et al., 2014); (2) Bias and contraction characterization: we quantify the deviation between classical push-forwards and amplitude-encoded marginals, showing how cross-terms vanish on permutation-like rows and how reversible gaps contract total-variation errors over steps; and (3) Empirical evaluation on hubbed graphs: across depths $d \in \{4,8,16\}$ and matched horizons, we observe high quantum-classical agreement in the complete dataset with systematic improvements after hub pruning: removing black reduces TVD/KL at shallow depth while preserving near-unit fidelity, and removing both hubs further regularizes the spectrum and yields numerical identity in our smallest instances.

## 2. Literature Review

Quantum algorithms have approached DTMCs primarily via quantum walks. In Szegedy's bipartite construction, a reversible chain is encoded by a unitary whose spectrum mirrors that of the classical transition operator, yielding quadratic improvements in search and hitting times, the canonical "$\sqrt{\delta\varepsilon}$" phenomenon, and, for ergodic chains with spectral gap $\gamma$, quantum hitting (and often mixing) times that scale as $O\left(\frac{1}{\sqrt{\gamma}}\right)$ rather than $O\left(\frac{1}{\gamma}\right)$ (Szegedy, 2004; Magniez et al., 2012). Subsequent work has analyzed continuous-time walks, quantum mixing, and fast-forwarding of Markov dynamics, typically targeting the same quadratic gains and clarifying the conditions under which they materialize (Apers & Sarlette, 2019; Li & Shang, 2023; Chakraborty et al., 2020; Childs & Goldstone, 2004; Krovi et al., 2010; Portugal, 2018; Ambainis, 2007; Venegas-Andraca, 2012).

Beyond walks, block-encoding and linear combinations of unitaries (LCU) embed stochastic matrices into unitaries, enabling polynomial and rational transformations of Markov operators via phase estimation and quantum singular-value transformation (Low & Chuang, 2017; Gilyén et al., 2019; Harrow et al., 2009). These primitives support one-step evolutions with oracle access and offer routes to bias state preparation toward stationary distributions or to approximate functions of the stationary measure (Wocjan & Abeyesinghe, 2008; Childs et al., 2017; Chakraborty et al., 2019; Aaronson, 2015; Berry et al., 2017). Across these results, hitting-time improvements hinge on spectral properties. In hub-dominated graphs, large imbalances of degree impair conditioning, while pruning or normalizing hubs serves as an explicit conditioning step that aligns with shallow-depth quantum implementations.

AE complements the above by accelerating Monte-Carlo type inference: estimating expectations to error $\varepsilon$ requires $O\left(\frac{1}{\varepsilon^2}\right)$ classical samples but only $O\left(\frac{1}{\varepsilon}\right)$ AE queries (Brassard et al., 2002; Montanaro, 2015). Noisy intermediate-scale quantum (NISQ)-oriented variants reduce circuit depth at some asymptotic cost (Giurgica-Tiron et al., 2020; Plekhanov et al., 2022; Suzuki et al., 2020; Maronese et al., 2023). In parallel, quantum routines for discrete distributions (divergences, entropy, and moments) are directly relevant to DTMC analysis. Parameterized circuits have been used to estimate total variation, KL, and related distances and to approximate covariance between discrete random variables under amplitude encoding (Peretz & Koren, 2024; Koren & Peretz, 2024). A complementary "black-box" circuit computes Shannon entropy to classical precision with constant-depth circuitry (Koren et al., 2023). Taken together, these tools form building blocks for evaluating transition measures and stationary-distribution functionals on quantum devices (Gilyén et al., 2019; Low & Chuang, 2017; Koren et al., 2023; Peretz & Koren, 2024; Wocjan & Abeyesinghe, 2008; Aaronson & Rall, 2020; Tanaka et al., 2020; Grinko et al., 2021; Suzuki & Yamamoto, 2021; Woerner & Egger, 2019; Koren & Peretz, 2024).

A central question is the comparative cost of DTMC estimation. Classical Monte Carlo scales as $O\left(\frac{1}{\varepsilon^2}\right)$ in precision and $O\left(\frac{1}{\gamma}\right)$ in spectral gap for mixing, whereas quantum methods promise at most quadratic improvements, with AE reducing precision scaling to $O\left(\frac{1}{\varepsilon}\right)$ and quantum walks suggesting $O\left(\frac{1}{\sqrt{\gamma}}\right)$ dependences (Montanaro, 2015; Magniez et al., 2012; Szegedy, 2004). In practice, these gains are conditional: state preparation and oracle access to transition probabilities may dominate, while circuit depth, T-gate counts, and device noise introduce overhead that can erase asymptotic advantages (Low & Chuang, 2017; Gilyén et al., 2019; Suzuki et al., 2020; Giurgica-Tiron et al., 2020; Plekhanov et al., 2022; Maronese et al., 2023). Hence, for stationary-distribution or one-step expectation estimation, quadratic speedups are the realistic ceiling and hinge on efficient data access and robust implementations (Montanaro, 2015; Magniez et al., 2012; Childs et al., 2019; Biamonte & Bergholm, 2017; Berry et al., 2015; Preskill, 2018; Gheorghiu et al., 2019).

Forecasting demand in the fashion industry presents unique challenges due to its rapid pace, seasonality, and the short lifecycle of products, which often lack sufficient historical sales data for reliable analysis (Koren & Peretz, 2025; Swaminathan & Venkitasubramony, 2023; Namin et al., 2022). Collections can change within weeks of runway presentations, and trend adoption typically follows a diffusion process influenced by social and cultural dynamics (often initiated by celebrities or influencers) before rising, peaking, and declining. This contributes to the industry's high degree of non-stationarity: consumer preferences shift frequently, demand is strongly seasonal (e.g., outerwear in winter), and external events such as viral moments, concerts, or cultural phenomena can cause sudden shifts in purchasing behavior (Lee & Workman, 2021; Machi et al., 2022; Ergin et al., 2022; Jin & Ryu, 2019). While some historical cyclicality can assist forecasting, prior research emphasizes that predictability varies by product attribute. Color trends, for instance, tend to follow more regular and modelable patterns than fabrics or silhouettes. Additionally, macroeconomic conditions can influence consumer color preferences, as seen with the emergence of "Depression Chic" aesthetics following the 2008 financial crisis (Silva et al., 2019; Ren et al., 2020; DuBreuil & Lu, 2020; Koh, 2019).

In response to this volatility, recent approaches have integrated heterogeneous data sources alongside AI-based forecasting models that capture multi-factor relationships and assign probabilistic confidence levels to trend predictions (Koren & Shnaiderman, 2023; Garcia, 2022; Koren et al., 2022). Improved forecasting not only supports more accurate commercial decision-making but also addresses sustainability concerns in a sector frequently criticized for overproduction and environmental impact (Koren et al., 2023). This context motivates a focus on short-horizon colour-transition dynamics, which offer a more stable and operationally actionable target for modeling than broader style or fabric shifts.

In fashion e-commerce, DTMCs naturally model user navigation and product-to-product transitions. Sequence-aware recommenders show that click/order trajectories carry predictive signal beyond i.i.d. (independence and identical distribution) assumptions (Kim et al., 2024). Hybrid architectures that fuse a Markov chain over purchases with an attribute graph capture temporal dynamics and improve recommendations (Zhang et al., 2021). Operationally, clickstream pipelines use higher-order DTMCs and adjacent sequence models to exploit short-term dependence (Scholz, 2016; Rendle et al., 2010; He et al., 2017; Wang et al., 2018; Quadrana et al., 2018; Sun et al., 2019; Xu et al., 2019).

Quantum estimators fit these pipelines at several touchpoints. Estimating transition probabilities between categories (e.g., dress to shoes) is a DTMC parameter-estimation task amenable to AE. Computing global rankings (PageRank-style) and diversity measures (e.g., entropy of category distributions) maps to quantum singular value transformation (QSVT)/quantum phase estimation (QPE)-based linear algebra routines and entropy circuits (Low & Chuang, 2017; Gilyén et al., 2019; Koren et al., 2023; Peretz & Koren, 2024; Sánchez-Burillo et al., 2012; Garnerone et al., 2012). Recent advances in covariance and distributional estimation (Koren & Peretz, 2024; Peretz & Koren, 2024) transfer to demand forecasting, in which cross-category and intertemporal dependencies must be quantified. Moreover, entropy-based methods help characterize uncertainty in preference dynamics, a recurring challenge under fast-changing assortments and sparse observations (Koren et al., 2023; Koren & Peretz, 2024).

Downstream forecast and inventory quality depend on robust preprocessing. Automated clustering and feature-selection pipelines and anomaly-detection procedures can stabilize inputs before modeling (Koren et al., 2025; Koren et al., 2023). When paired with classical ML, such preprocessing reduces noise and improves the reliability of sequence models. Anomaly-detection safeguards (Koren et al., 2023) are particularly relevant for shock events (e.g., sudden fads), during which rapid distributional shifts compromise DTMC stationarity. These considerations motivate hybrid architectures in which quantum primitives (AE, entropy, and divergence estimation) are embedded within classical learning stacks to yield resilient, scalable decision-support tools for fashion inventory management.

## 3. Method

This section presents a quantum procedure for simulating a DTMC over observable colour states. We specify notation and data structures, describe the gate-level primitives and their composition (state preparation, block-encoding, and oblivious amplitude amplification), provide the algorithmic flow for distributional updates and spectral diagnostics, and justify correctness via simple lemmas linking amplitude evolution to the classical push-forward.

**Table 1**. Notation used in this study.

| Symbol | Remarks |
|---|---|
| $n$ | Number of observable colours (states) |
| $N = 2^q$ | Padded dimension for amplitude encoding; $q = \lceil \log_2 n \rceil$ |
| $P \in \mathbb{R}^{n \times n}$ | Row-stochastic transition matrix ($P\mathbf{1} = \mathbf{1}$, $P_{ij} \geq 0$) |
| $p^{(t)} \in \Delta^{n-1}$ | Classical probabilities at step $t$ |
| $\lvert p \rangle$ | Amplitude-encoded distribution |
| $U_{\text{prep}}(p)$ | State-preparation unitary: $U_{\text{prep}}(p)\lvert 0^q \rangle = \lvert p \rangle$ |
| $U_P$ | $(\alpha, a)$ block-encoding of $P$ |
| $\alpha$ | Block-encoding scale, $\alpha \geq \lVert P \rVert_2$ |
| $R_0$ | Reflection operator |
| OAA | Oblivious amplitude amplification using $U_P$ and $R_0$ |
| $W(P)$ | Szegedy walk operator for spectral queries |

3.1. *Preliminaries*

Let $S = \{1, \ldots, n\}$ denote the set of colours (we pad to $N = 2^q$ with $q = \lceil \log_2 n \rceil$ if needed). The DTMC is specified by a row-stochastic matrix $P \in \mathbb{R}^{n \times n}$ such that $P_{ij} \geq 0$. Given an initial row distribution $p^{(0)} \in \Delta^{n-1}$, the classical $t$-step law is $p^{(t)} = p^{(0)} P^t$. For a distribution $p = (p_1, \ldots, p_n)$, we use amplitude encoding on $q$ system qubits:

$$\lvert p \rangle = \sum_{i=1}^{n} \sqrt{p_i} \, \lvert i \rangle \in \mathbb{C}^N$$

so that projective measurement in the computational basis samples $X \sim Ber(p)$.

We support two compilation paths:
1. Sparse-row oracle access: When each row $i$ has at most $s \ll n$ nonzero values, we assume
$$O_v: |i\rangle|j\rangle|0\rangle \mapsto |i\rangle|j\rangle|P_{ij}\rangle$$
$$O_s: |i\rangle|k\rangle \mapsto |i\rangle|j_k(i)\rangle$$
where $j_k(i)$ is the index of the $k^{\text{th}}$ nonzero value in row $i$.
2. Nonnegative LCU factorization: For dense or structured $P$ and $\lambda_i \geq 0$, we write
$$P = \sum_{i=1}^{L} \lambda_i A_i$$

such that $\sum_{i=1}^{L} \lambda_i = \alpha$, and $A_i$ is unitary and efficiently controlled. This preserves stochasticity in expectation and allows uniform treatment of sparse and dense regimes via the same block-encoding interface.

3.2. *Quantum Logic and Flow*

Let $p \in \Delta^{n-1}$ be a probability vector over the $n$ active colours, embedded in a $q$-qubit register ($N = 2^q \geq n$). A binary preparation tree induces cumulative weights $c_0^{(k)}, c_1^{(k)} \in [0,1]$ at node $k$, with $c_0^{(k)} + c_1^{(k)} = 1$. Using multiplexed rotations,
$$R_y(2\theta_k)|0\rangle = \sqrt{c_0^{(k)}} |0\rangle + \sqrt{c_1^{(k)}} |1\rangle$$
such that $\theta_k = \arccos\sqrt{c_0^{(k)}}$. Composing these controlled rotations along the tree yields
$$U_p(p)|0^q\rangle = |p\rangle = \sum_{j \in S} \sqrt{p_j} |j\rangle$$
where $S \subseteq \{0, \ldots, N-1\}$ indexes the active colours. If $n < N$, we set amplitudes to zero. We denote preparation inaccuracy by $\varepsilon_p$ in total variation when mapped to measurement statistics.

Let $p^{(t)} \in \Delta^{n-1}$ as the amplitude state and prepare $|p^{(t)}\rangle = \sum_{i=1}^{n} \sqrt{p_i^{(t)}} |i\rangle$, which yields the initial state
$$|\psi_0\rangle = |0^a\rangle \otimes |p^{(t)}\rangle$$
Let $U_P$ be an $(\alpha, a)$ block-encoding of $P$, i.e.,
$$(\langle 0^a| \otimes I) U_P (|0^a\rangle \otimes I) = \frac{P}{\alpha}$$
Applying $U_P$ promotes the system amplitudes according to
$$|\psi_1\rangle = U_P|\psi_0\rangle = |0^a\rangle \otimes \frac{P|p^{(t)}\rangle}{\alpha} + |\perp\rangle$$

Let $R_0 = (2|0^a\rangle\langle 0^a| - I) \otimes I$ and the Grover-like iterate $G = -U_P R_0 U_P^\dagger R_0$. On the two-dimensional subspace spanned by the good and orthogonal components, $G$ acts as a rotation by $2\theta$, where the success angle satisfies

$$\sin\theta = \|(\langle 0^a| \otimes I)U_P|\psi_0\rangle\| = \frac{\|P|p^{(t)}\rangle\|_2}{\alpha} \in (0,1]$$

After $r$ iterations of OAA and post selection on $|0^a\rangle$, the reduced system state is

$$\left|\tilde{p}_{AE}^{(t+1)}\right\rangle = \frac{P|p^{(t)}\rangle}{\|P|p^{(t)}\rangle\|_2}$$

To avoid over-rotation when $\theta$ is unknown, we employ a fixed-point OAA schedule achieving an angle error of at most $\varepsilon_{OA}$ with $O\left(\log\left(\frac{1}{\varepsilon_{OA}}\right)\right)$ reflections, preparing a state whose fidelity to $\left|\tilde{p}_{AE}^{(t+1)}\right\rangle$ is at least $1 - \varepsilon_{OA}^2$.

Let $\eta_j = \sum_i \sqrt{p_i^{(t)}} P_{ij}$. Measuring the system in the computational basis yields

$$P_{AE}[X_{t+1} = j] = \frac{\eta_j^2}{\sum_k \eta_k^2} = \frac{\left(\sum_i \sqrt{p_i^{(t)}} P_{ij}\right)^2}{\|P|p^{(t)}\rangle\|_2}$$

The classical push-forward $(p^{(t)}P)_j = \sum_i p_i^{(t)} P_{ij}$ is recovered when rows of $P$ are one-hot (permutations), making cross-terms vanish. In general, the quadratic form introduces cross-terms and thus constitutes a proxy for $p^{(t)}P$. AE is best used for spectral analysis and proxy marginals where rank ordering or coarse mass shifts suffice.

---

**Q-Colors(P, P0, T, M)**
- Allocate $q \leftarrow \lceil \log_2 n \rceil$ qubits and $a$ ancillas.
- $p_0 \leftarrow U_p(p)|0^q\rangle$
- For $t = 0$ to $T - 1$
  - $|\psi_0\rangle \leftarrow |0^a\rangle \otimes |p^{(t)}\rangle$
  - $U_P|\psi_0\rangle \leftarrow |0^a\rangle \otimes \frac{P|p^{(t)}\rangle}{\alpha} + |\bot\rangle$
  - $G \leftarrow -U_P R_0 U_P^\dagger R_0$
  - $\left|\tilde{p}_{AE}^{(t+1)}\right\rangle = \frac{P|p^{(t)}\rangle}{\|P|p^{(t)}\rangle\|_2}$
  - Measure $M$ shots
  - $\hat{p}_j^{(t+1)} \approx \left(\sum_i \sqrt{p_i^{(t)}} P_{ij}\right)^2$
- **Return** $\{\hat{p}^{(t)}\}_{t=0}^T$

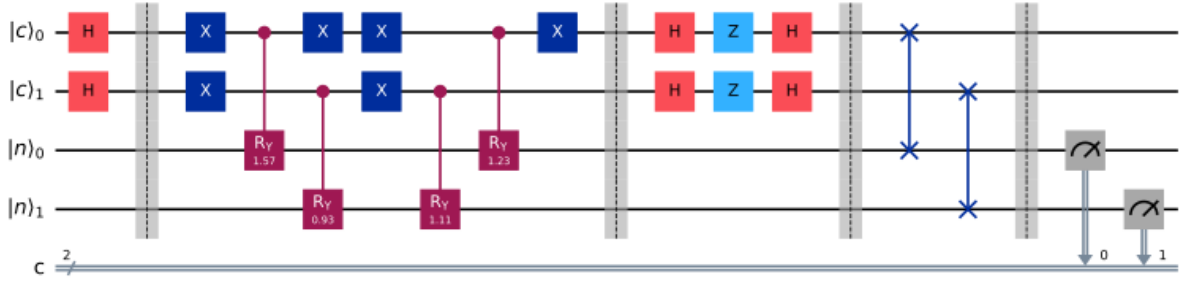

**Figure 1.** Quantum circuit for discrete-time Markov chain

### 3.3. Correctness

**Lemma 1.** *Let $U_P$ be an $(\alpha, a)$ block-encoding of $P$, $R_0 = (2|0^a\rangle\langle 0^a| - I) \otimes I$, and $G = -U_P R_0 U_P^\dagger R_0$. For any $|p^{(t)}\rangle$, define $\sin\theta = \frac{1}{\alpha}\|P|p^{(t)}\rangle\|_2 \in (0,1]$. After $r$ applications of $G$ to $|\psi_0\rangle = |0^a\rangle \otimes |p^{(t)}\rangle$, measurement of the ancilla with outcome $0^a$ leaves the system in state $|\tilde{p}_{AE}^{(t+1)}\rangle = \frac{P|p^{(t)}\rangle}{\|P|p^{(t)}\rangle\|_2}$ with success probability $\sin^2((2r+1)\theta)$. Fixed-point OAA with angle error at most $\varepsilon_{OA}$ prepares a state with fidelity at least $1 - \varepsilon_{OA}^2$ using $O(\log(\varepsilon_{OA}^{-1}))$ reflections.*

*Proof.* Given the block-encoding property $(\langle 0^a| \otimes I) U_P (|0^a\rangle \otimes I) = \frac{P}{\alpha}$, it follows that $U_P|\psi_0\rangle = \sin\theta |g\rangle + \cos\theta |b\rangle$, where $\sin\theta = \frac{1}{\alpha}\|P|p^{(t)}\rangle\|_2$ and $g = |0^a\rangle \otimes \frac{P|p^{(t)}\rangle}{\|P|p^{(t)}\rangle\|_2}$. The Grover operator $G = -U_P R_0 U_P^\dagger R_0$ acts as a rotation by $2\theta$ on $\{g, b\}$, hence:
$$G^r U_P |\psi_0\rangle = \sin((2r+1)\theta)|g\rangle + \cos((2r+1)\theta)|b\rangle$$
Post selecting ancilla $0^a$ yields $\frac{P|p^{(t)}\rangle}{\|P|p^{(t)}\rangle\|_2}$ with probability $\sin^2((2r+1)\theta)$. The fixed-point OAA variant achieve fidelity of at least $1 - \varepsilon_{OA}^2$ using $O(\log(\varepsilon_{OA}^{-1}))$ reflections.

**Lemma 2.** *Let $p^{(t)} \in \Delta^{n-1}$ and $\eta_j = \sum_i \sqrt{p_i^{(t)}} P_{ij}$. Then $\mathbb{P}_{AE}[X_{t+1} = j] = \frac{\eta_j^2}{\sum_k \eta_k^2}$. If each row of $P$ is one-hot (permutation/partition), then $\eta_j^2 = \sum_i p_i^{(t)} P_{ij}$ and $\mathbb{P}_{AE}[j] = (p^{(t)}P)_j$. In general:*
$$\eta_j^2 = \|v_j\|_2^2 + 2 \sum_{1 \le i < i' \le n} \sqrt{p_i^{(t)} p_{i'}^{(t)}} \cdot P_{ij} P_{i'j}$$

so the deviation from $(p^{(t)}P)_j$ is controlled by the normalized sum of cross-terms in the numerator and denominator.

*Proof.* For $p^{(t)} \in \Delta^{n-1}$ define $\eta_j = \sum_i \sqrt{p_i^{(t)}} P_{ij}$. The AE measurement distribution is $\Pr[X_{t+1} = j] = \frac{\eta_j^2}{\sum_k \eta_k^2}$. Expanding the square gives

$$\eta_j^2 = \sum_i p_i^{(t)} P_{ij}^2 + 2 \sum_{1 \le i < i' \le n} \sqrt{p_i^{(t)} p_{i'}^{(t)}} P_{ij} P_{i'j}$$

If each row of $P$ is one-hot, then $P_{ij}^2 = P_{ij}$ and all cross-terms vanish, yielding $\eta_j^2 = (p^{(t)}P)_j$. Because $\sum_j \eta_j^2 = 1$, this matches the classical distribution. In the general case, cross-terms remain, and the deviation is bounded precisely by their normalized contribution relative to $\sum_k \eta_k^2$. This establishes the bias characterization.

**Lemma 3.** *Assume reversibility with stationary $\pi$ and spectral gap $\gamma = 1 - \lambda_\star$, where $\lambda_\star = \max\{|\lambda_2|, |\lambda_n|\} < 1$ for the discriminant. Let $\mu^{(t)} = p^{(0)} P^t$ and $\hat{\nu}^{(t)}$ be the empirical AE marginal produced with per-step additive evolution error $\varepsilon = \varepsilon_p + \varepsilon_b + \varepsilon_{OA}$ and $M$ samples per step. Then for $t \ge 1$ and constant $c$,*

$$\mathbb{E}\left[\|\hat{\nu}^{(t)} - \mu^{(t)}\|_{TV}\right] \le \frac{\varepsilon}{\gamma} + \frac{c}{\sqrt{M}}$$

*Proof.* Let $\mu^{(t)} = p^{(0)} P^t$ denote the ideal distribution and $\nu^{(t)}$ the exact AE marginal after $t$ steps with per-step evolution error $\varepsilon_e$ (preparation, block-encoding, OAA). Total variation contracts under reversible $P$ with spectral gap $\gamma$:

$$\|\nu^{(t)} - \mu^{(t)}\|_{TV} \le (1-\gamma)\|\nu^{(t-1)} - \mu^{(t-1)}\|_{TV} + \varepsilon$$

Unrolling gives

$$\|\nu^{(t)} - \mu^{(t)}\|_{TV} \le \sum_{k=1}^{t} (1-\gamma)^{t-k} \varepsilon$$

Sampling $M$ times adds a stochastic deviation with expectation $O(1/\sqrt{M})$. Taking expectations over the sampling yields

$$\mathbb{E}\left[\|\hat{\nu}^{(t)} - \mu^{(t)}\|_{TV}\right] \le \sum_{k=1}^{t} (1-\gamma)^{t-k} \varepsilon + \frac{c}{\sqrt{M}}$$

Finally, summing the geometric series gives the uniform bound $\frac{\varepsilon}{\gamma} + \frac{c}{\sqrt{M}}$.

## 4. Empirical Study

We quantify how platform-level recommendation links among fashion products induce observable transitions between colour states and use these transitions to estimate DTMCs at the colour level.

We report construction details, state-space consolidation (from 91 tags to 27 states), model variants (with/without the primary colours: black and white), and evaluation metrics used to compare classical and quantum estimators.

4.1. *Data Collection and Processing*

Our analysis uses 85,659 products scraped from five large international online fashion retailers. To preserve partner anonymity, we do not disclose platform identities or raw HTML. For each product, we store metadata (price, text description, platform colour tag) and the platform-generated list of similar/recommended products exposed on the product page. We interpret recommendation links as observed navigation or preference transitions between product colours. Let $s$ be the colour of the viewed (source) product and $s'$ the colour of a recommended (target) product. Each recommendation edge contributes one count to $(s \to s')$. Aggregating over all pages yields matrix $C \in \mathbb{N}_0^{M \times M}$, such that $C_{ij} = \#\{\text{recs from colour } i \text{ to } j\}$.

Platform colour tags are normalized to WGSN categories (Appendix A), giving $M = 27$ distinct colours that define the state space $C$. We convert counts to row-stochastic probabilities via Laplace smoothing with $\beta = 0.1$:

$$\hat{P}_{ij} = \frac{C_{ij} + \beta}{\sum_{k=1}^{M}(C_{ik} + \beta)}, \qquad i = 1, \ldots, M$$

A preliminary inspection shows that black dominates recommendation flows, potentially masking structure among non-black colours. Because black and white are primary colours in fashion (Han ae al., 2022; Kodžoman et al., 2022; Carver, 2023), we analyze three DTMC variants built from the same graph:
- $M_1$ (full): $C$ with all $M = 27$ colours (Appendix A).
- $M_2$ (without black): induced subgraph on $C_b$ (26 states, without black). We recompute counts on the subgraph and renormalize each row:

$$\hat{P}_{ij}^{b} = \frac{C_{ij}^{b} + \beta}{\sum_{k \in C_b}(C_{ik}^{b} + \beta)}, \qquad i, j \in C_b$$

- $M_3$ (without black and white): induced subgraph on $C_{bw}$ (25 states, without black and white) with row-wise re-estimation:

$$\hat{P}_{ij}^{bw} = \frac{C_{ij}^{bw} + \beta}{\sum_{k \in C_{bw}}(C_{ik}^{bw} + \beta)}, \qquad i, j \in C_{bw}$$

The same assessment was performed to $M_3$ over all $M = 25$ colours excluding black and white. Re-estimating rows on the restricted support (rather than deleting black from $P_1$) preserves valid row-stochastic probabilities that reflect relative preferences among non-black colours in a hypothetical catalogue without black. For quantum compilation, we embed $P_1 \in \mathbb{R}^{27\times 27}$, $P_2 \in \mathbb{R}^{26\times 26}$, and $P_3 \in \mathbb{R}^{25\times 25}$ into $128 \times 128$ by zero-padding.

### 4.2. Evaluation Metrics

Let $P_c$ denote a classical reference distribution (or marginal) at a fixed horizon and $P_q$ its quantum estimate. We report the following:

- Total variation distance (TVD), which measures the maximum difference in probability mass between two distributions:

$$\text{TVD}(P_c, P_q) = \frac{1}{2}\sum_i \left| P_{c_i} - P_{q_i} \right|$$

- $L_2$ distance, which quantifies squared error magnitude:

$$\|P_c - P_q\|_2 = \sqrt{\sum_i \left(P_{c_i} - P_{q_i}\right)^2}$$

- Kullback-Leibler divergence (KL), which captures relative entropy and is well-defined on common support:

$$\text{KL}(P_c \| P_q) = \sum_i P_{c_i} \log \frac{P_{c_i}}{P_{q_i}}$$

- Fidelity ($F$), which measures geometric overlap of probability mass:

$$F(P_c, P_q) = \left(\sum_i \sqrt{P_{c_i} \cdot P_{q_i}}\right)^2$$

### 4.3. Experimental Procedure

- Each quantum configuration runs with 4096 shots. Classical baselines run the power method up to stated iteration/tolerance limits. Temporal evolution is simulated up to $T = 1000$ steps but summarized over the first 100 steps for efficiency.
- For a kernel of size $K \in \{25, 26, 27\}$, we choose $q \geq \lceil \log_2(K+1) \rceil$ and prepare amplitude states on $q$ qubits; zero-padded indices are excluded from metrics. The initial distribution is uniform on the active set.
- At each step, we apply the compiled Markov update: nonnegative LCU blocks realize the combination of unitary terms; for spectral queries we construct one Szegedy step and (optionally) a QPE probe. Measurement of the system register yields empirical marginals and repeated shots form $\hat{p}^{(t)}$.
- We use approximate state-prep with threshold $10^{-2}$, 4-step Trotterization of the walk, and reduced LCU width (32 terms). Depth is adapted from $\{1, 2, 4, 8, 16, 24, 32\}$ based on convergence and the 0.01 accuracy target.
- Each scenario records 4096 outcomes per horizon; TVD, $L_2$, KL, and fidelity are computed for $t \in \{1, 2, 3, 5, 10, 20\}$.

- Quantum marginals are compared against the classical reference at matched horizons. For stationary behaviour, we monitor convergence gaps versus the power method and report iteration and time to tolerance.

### 4.4. *Demonstration*

We illustrate one distributional update on a two-qubit system ($n = 4$ colours, padded to $N = 4 = 2^2$). Let the computational basis $\{|00\rangle, |01\rangle, |10\rangle, |11\rangle\}$ encode colours $\{a, b, c, d\}$ in that order. We take an explicit initial probability

$$\Pr(a), \Pr(b), \Pr(c), \Pr(d) = (0.5, 0.25, 0.125, 0.125)$$

and a row-stochastic transition is $P = \lambda \Pi_0 + (1 - \lambda) \Pi_1$, where $\lambda = 0.7$, $\Pi_0 = I_4$, and $\Pi_1 = I \otimes X$ (bit-flip on the least-significant qubit), i.e., each row mixes "stay" with probability $\lambda$ and "toggle last bit" with probability $1 - \lambda$.

We use the standard two-level rotation tree. Write $\mathbf{p} = [a, b, c, d]$ as two blocks: $P(a) + P(b)$ for the upper half $\{|00\rangle, |01\rangle\}$ and $P(c) + P(d)$ for the lower half $\{|10\rangle, |11\rangle\}$. Define

$$\theta_{\text{top}} = 2\arctan\sqrt{\frac{P(c) + P(d)}{P(a) + P(b)}} = 2\arctan\sqrt{\frac{0.25}{0.75}} = 1.0472 \text{ rad } (\approx 60°)$$

Apply $R_y(\theta_{\text{top}})$ to qubit $q_1$ (the most significant qubit). Conditioned on $q_1 = 0$ (upper block), apply $R_y(\theta_0)$ on $q_0$ with

$$\theta_0 = 2\arctan\sqrt{\frac{P(b)}{P(a)}} = 2\arctan\sqrt{0.5} = 1.2310 \text{ radians}$$

Conditioned on $q_1 = 1$ (lower block), apply $R_y(\theta_1)$ to $q_0$ with

$$\theta_1 = 2\arctan\sqrt{\frac{P(d)}{P(c)}} = 2\arctan 1 = \frac{\pi}{2}$$

Acting on $|00\rangle$, these three rotations prepare

$$|\psi(\mathbf{p})\rangle = (\sqrt{0.5}, \sqrt{0.25}, \sqrt{0.125}, \sqrt{0.125}) \cdot \begin{bmatrix} |00\rangle \\ |01\rangle \\ |10\rangle \\ |11\rangle \end{bmatrix} =$$

$$0.7071 |00\rangle + 0.5000 |01\rangle + 0.3536 |10\rangle + 0.3536 |11\rangle$$

Then $(I \otimes X)|\psi\rangle$ swaps the lower bit:

$$0.5000 |00\rangle + 0.7071 |01\rangle + 0.3536 |10\rangle + 0.3536 |11\rangle$$

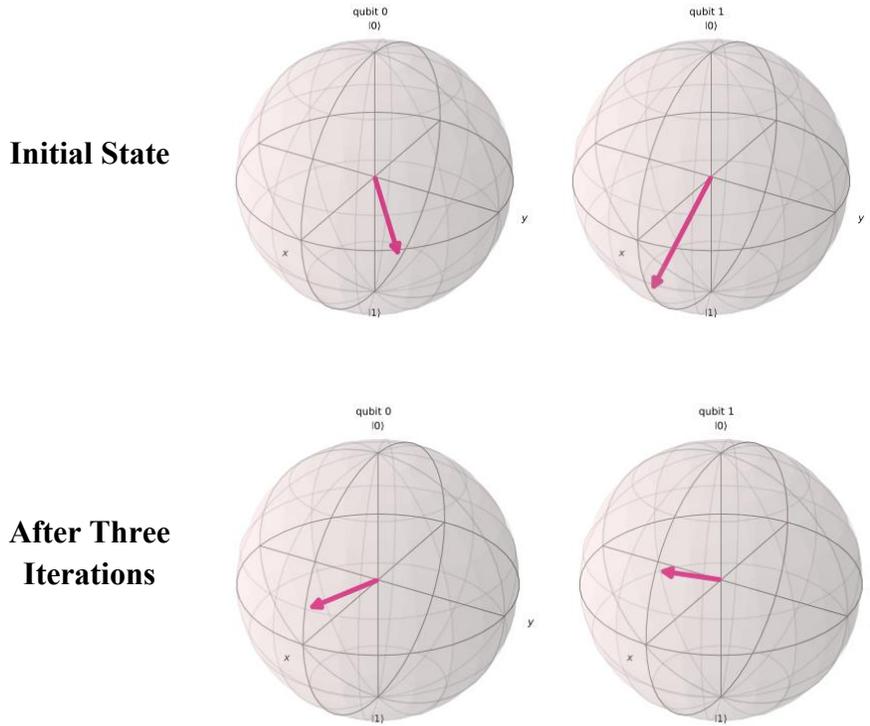

**Figure 2.** Qubits before and after application of quantum gates

Before computation, the (unnormalized) system amplitudes on ancilla $a = 0$ are
$$\beta_i \approx \sqrt{\lambda}\, \alpha_i + \sqrt{1-\lambda}\, \tilde{\alpha}_i, \quad \sqrt{\lambda} = 0.83666, \quad \sqrt{1-\lambda} = 0.54772$$
Numerically,
$$\beta_i \approx (0.8655,\ 0.8056,\ 0.4887,\ 0.4887)$$
To form a probability distribution, we compute squared magnitudes and divide by their sum
$$|\beta_{00}|^2 = 0.8655^2 = 0.7491$$
$$|\beta_{01}|^2 = 0.8056^2 = 0.6490$$
$$|\beta_{10}|^2 = 0.4887^2 = 0.2388$$
$$|\beta_{11}|^2 = 0.4887^2 = 0.2388$$
Therefore, the normalized post-selected distribution is
$$P_q = \frac{1}{\sum_j |\beta_j|^2}(0.7491,\ 0.6490,\ 0.2388,\ 0.2388) = (0.399, 0.346, 0.127, 0.127)$$
For our $P = \lambda I + (1-\lambda)(I \otimes X)$, the classical equivalent:
$$(\lambda a + (1-\lambda)b,\ (1-\lambda)a + \lambda b,\ \lambda c + (1-\lambda)d,\ (1-\lambda)c + \lambda d)$$
$$= (0.425,\ 0.325,\ 0.125,\ 0.125)$$
Classical computing yielded the probability vector $(0.425, 0.325, 0.125, 0.125)$, whereas the quantum output is $(0.399, 0.346, 0.127, 0.127)$. The error between the classical and quantum computations:

$$\frac{|0.425 - 0.399| + |0.325 - 0.346| + 2 \cdot |0.125 - 0.127|}{4} = 0.012$$

## 5. Results

This section synthesizes the empirical evidence obtained from our quantum compilation of colour-transition Markov chains against their classical references. We evaluate quantum compilations of colour-transition DTMCs against classical references under three taxonomies and evaluation metrics described in Section 4.

### 5.1. *Main findings*

Figure 3 displays induced colour-level DTMCs for the three chains. Given that $P \in [0,1]^{M \times M}$ denote the one-step transition matrix, for colour $c$, we define out-flow/in-flow centralities and a coverage ratio by $H_{\text{out}}(c) = \sum_j P_{cj}$, $H_{\text{in}}(c) = \sum_i P_{ic}$, and $\Gamma(c) = \frac{|\{j : P_{cj} > 0\}|}{M-1}$. Empirically, in $M_1$, black has $H_{\text{out}}(\text{black}) \approx 1$ and $\Gamma(\text{black}) \approx 1$, behaving as a near-universal emitter, while white shows a similar but milder pattern. Such hubs short-circuit the graph, masking medium-probability chromatic relations. Removing black ($M_2$) reduces hub dominance and increases the effective use of non-hub edges. Removing both black and white ($M_3$) exposes secondary structures (e.g., charcoal → navy, beige → gold). Hub removal redistributes mass in the fundamental matrix $Z = (I - P + \mathbf{1}\pi^T)^{-1}$ by decreasing $\|P_c\|$ for hub $c$, which increases pairwise effective resistances among remaining nodes and improves identifiability of off-hub transitions.

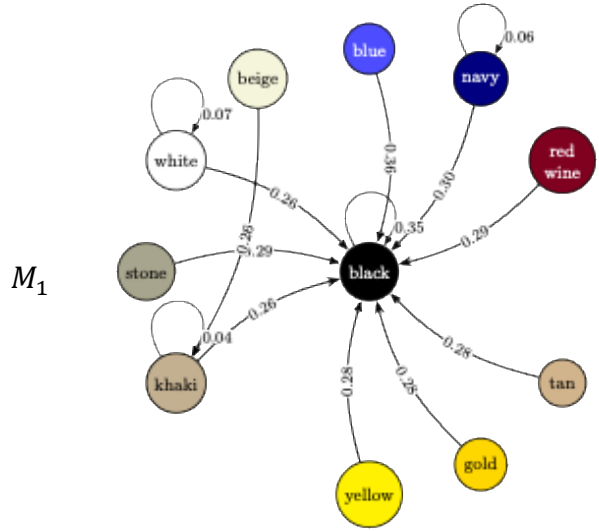

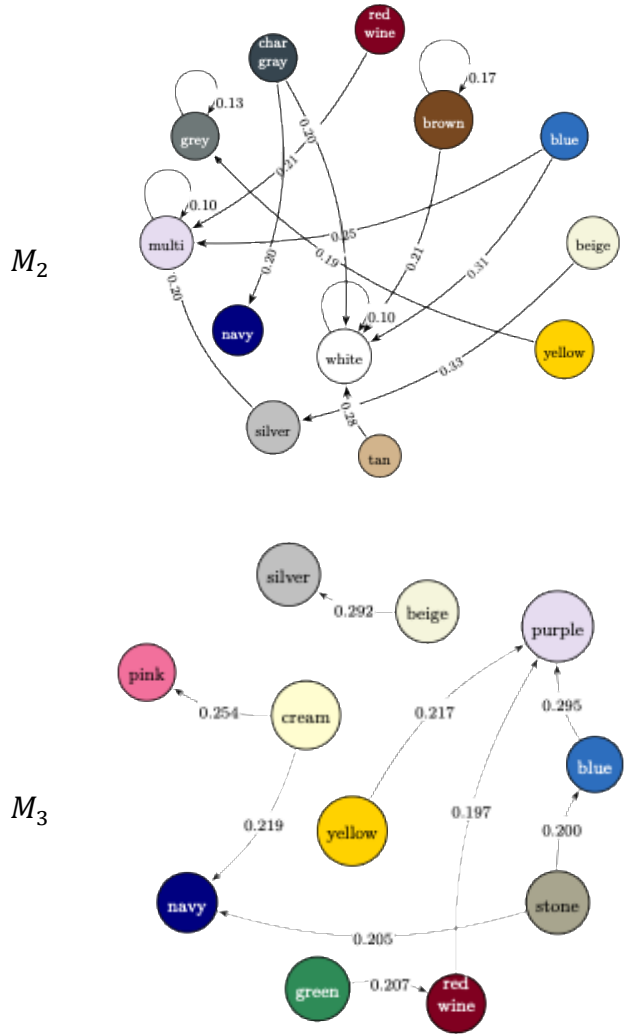

**Figure 3.** State diagram of $M_1, M_2,$ and $M_3$ DTMCs

### 5.2. Comparison across circuit depth

Table 2 summarizes accuracy vs. depth. All regimes are highly accurate: even in the most connected case ($M_1$), TVD $\leq 0.007$ and $F \geq 0.989$. Hub removal lowers divergence at shallow depth: at depth 4, $M_2$ results in half the TVD of $M_1$ (0.002 vs. 0.004) and reduces KL by $\approx 55\%$. With both hubs removed ($M_3$), the subgraph is sparser and more regular, and the circuit attains numerical identity with the classical chain (TVD $= 0$, $F = 1$), consistent with amplitude encoders on small, well-separated supports. Thus, $M_3$ is not included in Table 2.

**Table 2.** Quantum-classical agreement over circuit depth

| Model | Depth | $t$ | TVD | $L_2$ | KL | Fidelity |
|---|---|---|---|---|---|---|
| $M_1$ | 4 | 20 | 0.0034 | 0.0018 | $1.7\times 10^{-4}$ | 0.9888 |
| | | 50 | 0.0042 | 0.0019 | $2.1\times 10^{-4}$ | 0.9898 |
| | | 100 | 0.0038 | 0.0018 | $2.3\times 10^{-4}$ | 0.9917 |
| | 8 | 20 | 0.0042 | 0.0019 | $1.8\times 10^{-4}$ | 0.9901 |
| | | 50 | 0.0052 | 0.0026 | $2.9\times 10^{-4}$ | 0.9942 |
| | | 100 | 0.0054 | 0.0026 | $2.6\times 10^{-4}$ | 0.9999 |
| | 16 | 20 | 0.0066 | 0.0030 | $4.7\times 10^{-4}$ | 0.9998 |
| | | 50 | 0.0072 | 0.0033 | $4.5\times 10^{-4}$ | 0.9998 |
| | | 100 | 0.0077 | 0.0036 | $6.2\times 10^{-4}$ | 0.9997 |
| $M_2$ | 4 | 20 | 0.0027 | 0.0015 | $1.4\times 10^{-4}$ | 0.9988 |
| | | 50 | 0.0027 | 0.0014 | $9.6\times 10^{-5}$ | 0.9994 |
| | | 100 | 0.0034 | 0.0016 | $7.0\times 10^{-5}$ | 0.9999 |
| | 8 | 20 | 0.0045 | 0.0025 | $1.4\times 10^{-4}$ | 0.9995 |
| | | 50 | 0.0043 | 0.0021 | $1.3\times 10^{-4}$ | 0.9999 |
| | | 100 | 0.0039 | 0.0020 | $1.5\times 10^{-4}$ | 0.9999 |
| | 16 | 20 | 0.0063 | 0.0028 | $2.8\times 10^{-4}$ | 0.9999 |
| | | 50 | 0.0083 | 0.0038 | $5.4\times 10^{-4}$ | 0.9997 |
| | | 100 | 0.0062 | 0.0030 | $5.1\times 10^{-4}$ | 0.9998 |

Because raw KL values are very small and vary multiplicatively, we analyze log(KL) across depths and matched horizons. Figure 4 shows mild increases in log(KL) with depth for both models, but with log(KL) for $M_2$ remaining consistently below that for $M_1$ at each depth. Confidence bands overlap at depth 16, indicative of diminishing error differences at moderate depth ($M_3$ exhibits near-zero KL across settings and is omitted from this analysis).

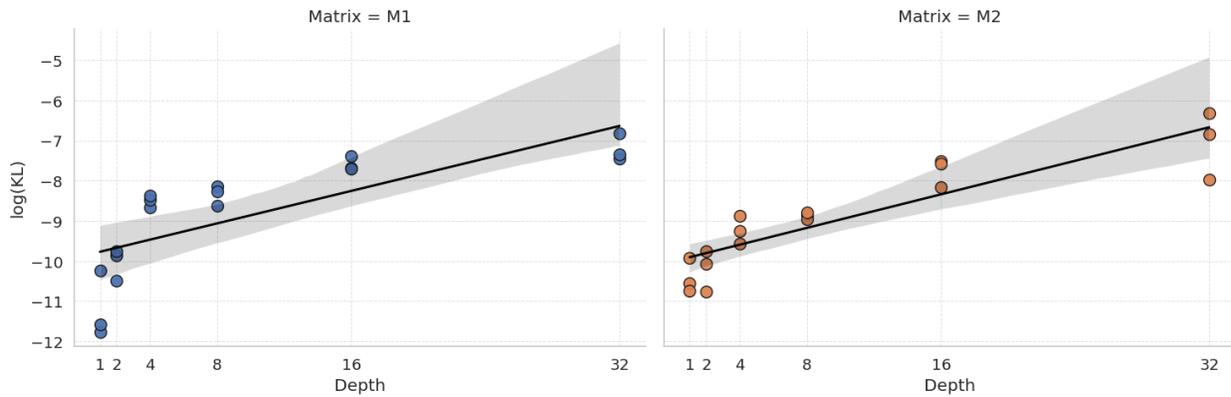

**Figure 4.** Comparison of log(KL) by depth between $M_1$ and $M_2$ DTMCs

We compare the number of iterations needed to reach tolerance against average stationary discrepancies, highlighting that removing black improves mixing efficiency and short-horizon accuracy. For the full catalogue with black included ($M_1$), convergence required about 35 iterations with an average runtime of 1.42 seconds, yielding a mean TVD of 0.061 and fidelity of 0.95. By contrast, removing black ($M_2$) reduced the computational effort to 18 iterations and 0.97 seconds runtime, with lower stationary discrepancy (average TVD 0.048) and slightly higher fidelity (0.96). Results for $M_3$, in which both black and white were pruned from the chain, were not reported because the removal of both hubs altered the structure of the stationary distribution.

Behaviourally, black and white act as universal matchers in styling; in the chain, they manifest as high-degree emitters with near-complete coverage Γ. Their presence concentrates stationary mass exchanges through hubs, obscuring medium-probability chromatic flows that are informative for downstream tasks (e.g., colour-aware policy learning, clustering). Removing black ($M_2$) and then white ($M_3$) decreases $\max_c \Gamma(c)$ (hub dominance), elevates visibility of non-hub transitions (e.g., charcoal to navy, beige to gold), and simplifies proposal distributions inside the quantum estimator, improving shallow-depth metrics and convergence.

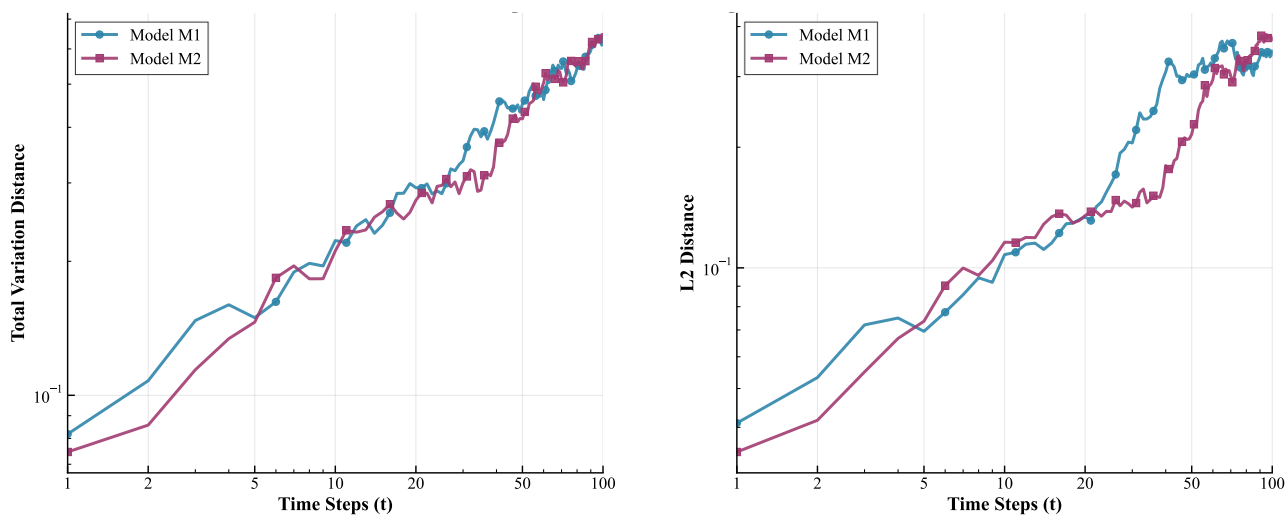

**Figure 5.** TVD and $L_2$ convergence comparison over time steps

Figure 5 compares the temporal evolution of the TVD and the $L_2$ distance between classical and quantum probability vectors across 100 simulation steps for the full ($M_1$) and hub-pruned ($M_2$) chains. In both metrics, divergence grows monotonically with time as the number of circuit updates and accumulated noise increase. However, the rate and magnitude of growth differ systematically between the two models. For early and intermediate horizons ($t \leq 30$), $M_2$ consistently maintains lower discrepancies, indicating that hub pruning improves short-term alignment between the quantum and classical trajectories. The gap narrows only at long horizons, when statistical fluctuations dominate, suggesting that the hub-induced conditioning effects are most pronounced in the early transient phase.

The logarithmic scaling in both plots emphasizes that the divergence growth follows an approximately exponential trend up to $t \approx 50$, after which both curves saturate. The slower ascent of $M_2$ reflects improved conditioning of the block-encoded operator: because amplitude concentration is reduced around a small set of high-probability transitions, the effective spectral gap widens, and the quantum propagation better captures medium-probability transitions without overshooting. In contrast, the full chain ($M_1$) exhibits steeper initial growth, consistent with faster amplification of dominant hub states and underrepresentation of weaker links. Together, these dynamics confirm that hub removal stabilizes early-stage quantum evolution and improves fidelity in limited-depth regimes.

## 6. Discussion

This work presented a quantum compilation framework for estimating colour-transition dynamics in fashion e-commerce. Its innovative aspect lies in the adaptation of amplitude-encoded Markov chains to a domain with pronounced hub states, in which colours such as black and white dominate recommendation flows. By combining block-encoded updates with amplitude amplification, the method allows shallow-depth circuits to recover key distributional features, while pruning of hub states highlights structural dependencies that are otherwise obscured. In doing so, the study demonstrates how graph topology interacts with quantum compilation and affects estimation quality at practical circuit depths. These observations are consistent with the broader theory of quantum walks and amplitude-based estimation, which links performance to the spectral properties and conditioning of the underlying Markov operator (Szegedy, 2004; Magniez et al., 2012; Childs & Goldstone, 2004; Apers & Sarlette, 2019; Chakraborty et al., 2020). The main findings can be summarized as follows:

- **Hub dominance and identifiability**. Colour-transition graphs exhibit strong hub dominance: black (and to a lesser extent white) act as universal matchers, concentrating probability mass and masking medium-probability paths. In the full chain, this produces excellent quantum-classical agreement overall but reduces the identifiability of non-hub relationships. From a quantum-walk perspective, hubbing compresses the effective spectral spread that walk-based constructions exploit. From an amplitude-estimation perspective, it increases amplitude concentration on a few outcomes, raising finite-shot and finite-depth sensitivity for small-mass events (Szegedy, 2004; Magniez et al., 2012). Both mechanisms align with theory predicting that quantum gains are most visible when spectra are well separated and success amplitudes are not excessively skewed (Apers & Sarlette, 2019; Chakraborty et al., 2020).
- **Effect of pruning hubs**. Removing black and then both black and white progressively increases the effective use of non-hub edges, improves identifiability of off-hub transitions, and enhances shallow-depth accuracy (yielding lower TVD/KL with comparable or higher fidelity). Conceptually, pruning hubs enlarges the effective spectral gap on the residual subgraph and surfaces secondary structures (e.g., charcoal to navy, beige to gold) that would otherwise be masked. This is consistent with quantum-walk analyses in which mixing/hitting advantages scale with the square root of classical gap parameters, and with QSVT-style reasoning indicating that better-conditioned singular values yield more stable low-degree polynomial transforms under depth constraints (Szegedy, 2004; Magniez et al., 2012; Low & Chuang, 2017; Low & Chuang, 2019).

- **Shallow-depth agreement**. With modest depth, amplitude-encoded marginals reproduced classical references at high fidelity, and the gains from hub removal were most visible when noise and finite-depth bias were most consequential. This matches practical findings on low-depth amplitude-estimation variants and fixed-point amplification schedules: in skewed distributions, small-probability estimates suffer from sampling and over-rotation error. Balancing the mass (by hub pruning) improves convergence of AE-like estimators at realistic shot budgets (Yoder, Low & Chuang, 2014; Suzuki et al., 2020).

In sequence-aware recommendation, short-horizon transitions and attribute-level state spaces (e.g., colours) are central features. Our findings suggest that a hybrid procedure is most effective: treat hub colours as control variates or degree-normalized nodes to regularize the transition operator before quantum estimation, and then use quantum subroutines for precision-critical statistics (small-mass flows, divergences, entropy, covariance) that drive ranking diversity or cold-start exploration. This is compatible with PageRank-style/spectral components and with contemporary sequence models that leverage Markovian structure most within richer neural stacks (Koren & Bell, 2015; He et al., 2017; Ludewig et al., 2021; Pedro et al., 2024).

Several constraints temper our conclusions. The experiments used small qubit registers with zero-padding overhead, approximate state preparation, and finite-depth block encodings, each introducing controllable but non-negligible bias. Shot noise and device-oriented simplifications (reduced LCU width, depth-limited OAA schedules) disproportionately affect small probabilities and accumulate across steps—precisely why hub-pruned chains (with more even mass) show stronger shallow-depth agreement. On the data side, transition counts derive from platform-exposed recommendations rather than from full user trajectories and thus reflect embedded business logic and seasonality. Finally, the first-order stationarity assumption abstracts away temporal context and high-order dependencies commonly exploited by production recommenders. Integrating quantum estimates into such pipelines will require explicit treatment of context and higher-order effects (Rendle et al., 2010; Quadrana et al., 2018).

Future work may address the following directions:

- **Scalability and conditioning**. Pursue sparse-oracle designs and QSVT-based compilations that reduce block-encoding scale, combined with degree-normalized discriminants or lazy-walk transforms to regularize hubbed graphs. This should reduce depth and improve numerical stability over larger colour sets (Low & Chuang, 2017; Apers & Sarlette, 2019).
- **Hybridization**. Embed quantum divergence/entropy/small-mass estimators as plug-ins within sequence-aware recommenders, reserving quantum calls for precision-critical subroutines while delegating representation learning/personalization to classical models. This aligns with evidence that hybrid stacks capture temporal signal while benefiting from variance-reduced primitives at the edges (He et al., 2017; Ludewig et al., 2021).
- **Algorithmic robustness**. Explore error-mitigated amplitude-estimation variants, shadow-tomography-style estimators for distributional metrics, and learning-based state preparation that amortizes compilation across similar assortments. These directions target the known bottlenecks—state prep, oracle width, and depth budgets—highlighted in both the QSVT/AE literature and our empirical sensitivity analysis (Yoder, Low & Chuang, 2014; Suzuki et al., 2020; Low & Chuang, 2019).

Together, these strands should extend the encouraging small-system results observed here to larger catalogues and more volatile fashion cycles, while preserving the interpretability of colour-level Markov structure that motivates the application.

**Appendix A.** Colour mapping according to the WGSN code.

| Name | Colour | WGSN code | WGSN name |
|---|---|---|---|
| asphalt | 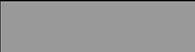 | 158-56-00 | zinc |
| auburn | 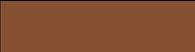 | 024-37-20 | nutshell |
| bayberry | 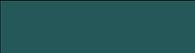 | 092-38-21 | verdigris |
| beige | 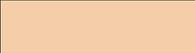 | 036-82-16 | vanilla cake |
| berry | 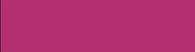 | 006-39-31 | raspberry pink |
| black | 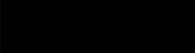 | 153-19-00 | black |
| blue | 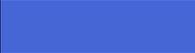 | 117- 47- 13 | elemental blue |
| brown | 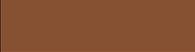 | 024-37-20 | nutshell |
| burgundy | 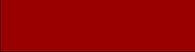 | 011-27-26 | bloodstone |
| camel | 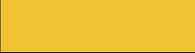 | 028-59-26 | sundial |
| cappuccino | 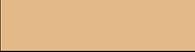 | 031-63-17 | parchment |
| chalk | 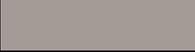 | 035-73-04 | sustained grey |
| champagne | 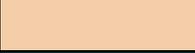 | 036-82-16 | vanilla cake |
| charcoal | 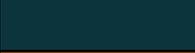 | 120-22-18 | midnight blue |
| chocolate | 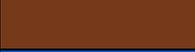 | 019-27-17 | sepia |
| cobalt | 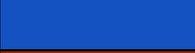 | 120-28-32 | galactic cobalt |
| cognac | 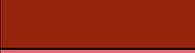 | 017-43-20 | terracotta |
| coral | 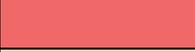 | 009-58-31 | sunset coral |
| cream | 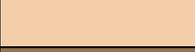 | 036-82-16 | vanilla cake |
| goat | 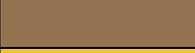 | 015-33-25 | intense rust |
| gold | 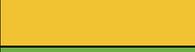 | 028-59-26 | sundial |
| green | 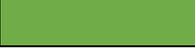 | 065-68-24 | apple mint |
| greige | 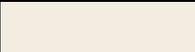 | 030-69-10 | oat milk |
| grey | 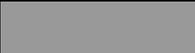 | 158-56-00 | zinc |
| indigo | 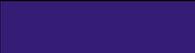 | 151-22-09 | midnight plum |
| kangaroo | 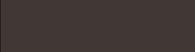 | 017-23-07 | dark oak |
| khaki | 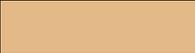 | 031-63-17 | parchment |
| lilac | 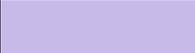 | 134-67-16 | digital lavender |
| marshmallow | 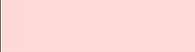 | 147-70-20 | fondant pink |
| mint | 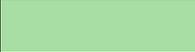 | 065-80-23 | neo mint |
| multi-coloured | | - | - |
| mustard | 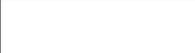 | 034-70-33 | mellow yellow |
| navy | 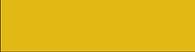 | 122-25-24 | lazuli blue |
| neutral | 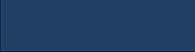 | 030-69-10 | oat milk |
| oatmeal | 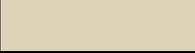 | 030-69-10 | oat milk |
| off-white | 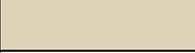 | 037-93-00 | optic white |
| olive | 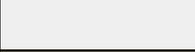 | 044-52-13 | olive oil |
| orange | 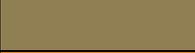 | *030-67-34* | mango sorbet |
| peat | 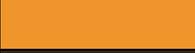 | 017-23-07 | dark oak |
| pink | 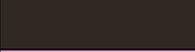 | 147-64-24 | wild rose |
| purple | 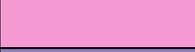 | 134-52-27 | jacaranda flower |
| putty | 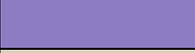 | 030-69-10 | oat milk |
| red | 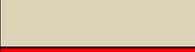 | 013-43-37 | red glow |
| rust | 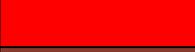 | 015-33-25 | intense rust |
| sage | 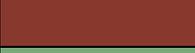 | 072-45-06 | sage leaf |
| sand | 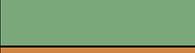 | 028-59-26 | sundial |
| sepia | 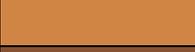 | 024-37-20 | nutshell |
| silver | 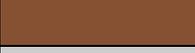 | 157-72-02 | pearl grey |
| superfruit | 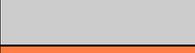 | 024-65-27 | apricot crush |
| stone | 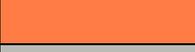 | 035-73-04 | sustained grey |
| tan | 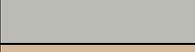 | 021-80-08 | transcendent pink |
| tidepool | 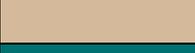 | 092-38-21 | verdigris |
| tobacco | 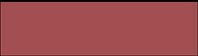 | 010-42-20 | astro dust |
| tradewinds | 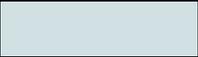 | 105-75-12 | cirrus blue |
| turquoise | 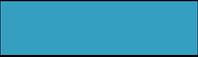 | 098-59-30 | A.I. aqua |
| trooper | 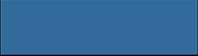 | 114-57-24 | tranquil blue |
| vanilla | 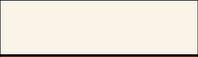 | 038-96-20 | wooden ruler |
| walnut | 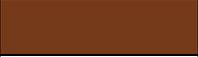 | 019-27-17 | sepia |
| white | 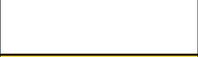 | - | - |
| yellow | 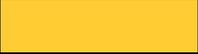 | 037-82-32 | ray flower |